\documentclass[12pt]{article}
\input epsf.sty
\usepackage{mathrsfs}
\usepackage{amsmath}
\usepackage{mathrsfs}
\usepackage{amssymb}
\usepackage{color}
\usepackage{psfrag}
\usepackage{graphicx}
\usepackage{subfigure}
\usepackage{rotating}

\newcommand{\bea}{\begin{eqnarray}}
\newcommand{\eea}{\end{eqnarray}}
\newcommand{\be}{\begin{equation}}
\newcommand{\ee}{\end{equation}}
\newcommand{\vs}[1]{\vspace{#1 mm}}

\newcommand{\dsl}{\pa \kern-0.5em /}

\newcommand{\half}{\frac{1}{2}}
\newcommand{\pa}{\partial}

\newcommand{\nn}{\nonumber\\}

\begin{document}
\topmargin 0pt
\oddsidemargin 0mm

\begin{flushright}



\end{flushright}

\vspace{2mm}

\begin{center}
{\Large \bf Intersecting D-branes and Lifshitz-like space-time}

\vs{10}

{Parijat Dey\footnote{E-mail: parijat.dey@saha.ac.in} and 
Shibaji Roy\footnote{E-mail: shibaji.roy@saha.ac.in}}

 \vspace{4mm}

{\em

 Saha Institute of Nuclear Physics,
 1/AF Bidhannagar, Calcutta-700 064, India\\}

\end{center}

\vs{10}

\begin{abstract}
In a previous paper \cite{Dey:2012tg} we have shown how Lifshitz-like
space-times (space-times having Lifshitz scaling with hyperscaling 
violation) arise from 1/4 BPS, threshold F-D$p$ bound state solutions of 
type II string theories in the near horizon limit. In this paper we show 
that similar structures also arise from the near horizon limit of 1/4 BPS, 
threshold intersecting D-brane solutions of type II string theories. Some of 
these solutions are standard (D$p$-D($p+4$) for $p=0,\,2$) and some are 
non-standard (D$p$-D($p+2)$ for $p = 1,\,2,\,3$) including D2-D2$'$,
D3-D3$'$ and D4-D4$'$ solutions. 
The dilatons of these solutions in general run (except in D2-D4 and
D3-D3$'$ cases) and 
produce RG flows. We discuss the phase structures of these solutions. D2-D4
and D3-D3$'$ 
in the near horizon limit do not produce Lifshitz-like space-time, but 
give AdS$_3$ spaces.     
\end{abstract}

\newpage

\section{Introduction}

Lifshitz scaling symmetry, a nonrelativistic symmetry, arises as a possible 
symmetry in some condensed matter systems at the quantum critical point 
\cite{Sachdev,Sachdev:2012dq}. 
As the system at this point is strongly coupled,
it can be studied holographically by using the general idea of AdS/CFT 
correspondence \cite{Maldacena:1997re} if a gravity dual, which 
asymptotes to a space-time 
with a Lifshitz scaling symmetry, can be found for such systems. Indeed such 
metrics
were found in \cite{Kachru:2008yh} as solutions of pure gravity theory 
coupled to matter.
Inclusion of dilaton enlarges the domain of such scaling symmetry of the
metrics \cite{Gubser:2009qt,Goldstein:2009cv,Cadoni:2009xm,
Charmousis:2010zz,Perlmutter:2010qu,Bertoldi:2010ca,Iizuka:2011hg,
Berglund:2011cp}. The string embeddings of this class of metrics were obtained 
in \cite{Hartnoll:2009ns}.

A more general class of scaling metrics (i.e., metrics with
Lifshitz-like scaling, namely, a Lifshitz scaling with a dynamical
critical exponent $z$ and a hyperscaling violation exponent $\theta$)
in the infrared have been found \cite{Ogawa:2011bz} by
using a general scaling argument, the logarithmic violation of the 
entanglement entropy and the null energy condition. Holographically they 
represent compressible metallic states with hidden Fermi surface 
\cite{Huijse:2011ef}. The whole
class of such scaling metrics are obtained as solutions to pure gravity
theories coupled to both the dilaton and an abelian gauge field 
\cite{Charmousis:2010zz}. Aspects of holography and some string
theory embeddings of these class of metrics have been obtained in 
\cite{Ogawa:2011bz,Dong:2012se,Narayan:2012hk,Kim:2012nb,Singh:2012un}. 
In a previous paper \cite{Dey:2012tg} we have shown how such metrics arise 
from the near horizon
limit of some unusual 1/4 BPS, threshold F-D$p$ bound state solutions of type
II string theories. In this paper, we show that similar structures also arise
from the near horizon limit of intersecting D-brane solutions of type II 
string theories. Some of these solutions are the standard 1/4 BPS threshold
intersecting D$p$-D$(p+4)$ (with $p=0,\,2$) solutions of type IIA string
theory and some are non-standard 1/4 BPS threshold intersecting D$p$-D($p+2$)
(with $p=1,\,2,\,3$) including D2-D2$'$, D3-D3$'$ and D4-D4$'$ solutions of 
type II string theories.
The dilatons for all these solutions (except D2-D4 and D3-D3$'$) are 
non-constant and
therefore produce RG flows. We discuss the phase structures for these solutions
and find that the metrics in other phases also have similar scaling
structures. The near horizon metric of D2-D4 and D3-D3$'$ solutions do not 
have 
Lifshitz-like scaling symmetry, but have the structures of AdS$_3$-spaces in
both phases. 
  
This paper is organized as follows. In section 2, we discuss the
standard intersecting D$p$-D$(p+4)$ (for $p=0,\,2$) solutions of type IIA
string theory their 
near horizon limits, scaling structures and the phase structures. In 
section 3, we discuss the same for the non-standard intersecting D$p$-D$(p+2)$
(for $p=1,\,2,\,3$) solutions along with D2-D2$'$, D3-D3$'$ and D4-D4$'$ 
solutions of type II string theories. Then we conclude in section 4.

\section{D$p$-D$(p+4)$ and Lifshitz-like
metrics}

In this section we will show that the standard 1/4 BPS threshold intersecting
D$p$-D$(p+4)$ solutions of type II string theories in the near horizon limit 
yield Lifshitz-like metrics. For $p=1$, we know that D1-D5 solution of type
IIB string theory does not give Lifshitz-like space-time but gives 
AdS$_3$ $\times$ S$^3$ $\times$ E$^4$ in the near horizon limit. So, we will 
consider only $p=0,\,2$ in the following. The string metric and the other field
configurations for D$p$-D$(p+4)$ intersecting solutions have the forms (see
for example, \cite{Tseytlin:1997cs}),
\bea\label{dpdp4}
ds^2 &=& H_1^{\half}H_2^{\half} \left[H_1^{-1}H_2^{-1} \left(-dt^2 + 
\sum_{i=1}^p(dx^i)^2\right)+H_2^{-1}\sum_{j=p+1}^{p+4}(dx^j)^2
+dr^2+r^2d\Omega_{4-p}^2\right]\nn
e^{2\phi} &=& H_1^{\frac{3-p}{2}}H_2^{-\frac{p+1}{2}}\nn
A_{[p+1]} &=& \left(1-H_1^{-1}\right)dt\wedge dx^1 \wedge \ldots \wedge dx^p,
\quad A_{[p+5]}\,\,=\,\, \left(1-H_2^{-1}\right)dt\wedge dx^1 \wedge \ldots 
\wedge dx^{p+4}\nn
\eea  
where the two harmonic functions have the forms $H_{1,2} = 1 +
Q_{1,2}/r^{3-p}$. $Q_{1,2}$ are the charges associated with D$p$ and D$(p+4)$
branes. The D$p$ branes are along $x^1,\ldots,x^p$ and D$(p+4)$ branes are
along $x^1,\ldots,x^{p+4}$. $r$ is the transverse radial coordinate given by
$r = \sqrt{(x^{p+5})^2 + \cdots +(x^9)^2}$. Note from \eqref{dpdp4} that for 
both $p=0,\,2$, dilaton $\phi$
are not constant and we have put the string coupling $g_s=1$. $A_{[p+1]}$ and
$A_{[p+5]}$ are the RR form fields which couple to D$p$ branes and D$(p+4)$
branes respectively\footnote{Here and below the constant terms in the 
form-fields are added such that the solution is asymptotically flat. 
However, in the near horizon limit when we deal with asymptotically non-flat 
solutions, we ignore the constant terms in the form fields.}.

In the near horizon limit we approximate $H_{1,2} \approx Q_{1,2}/r^{3-p}$.
Substituting this in the metric in \eqref{dpdp4} and further making a
coordinate transformation $r \to 1/r$ we get,
\be\label{metricdpdp4}
ds^2 = \sqrt{Q_1Q_2}r^{1-p}\left[-\frac{dt^2}{Q_1Q_2r^{4-2p}} +
  \frac{\sum_{i=1}^p (dx^i)^2}{Q_1Q_2r^{4-2p}} + \frac{\sum_{j=p+1}^{p+4} 
(dx^j)^2}{Q_2r^{1-p}} + \frac{dr^2}{r^2} + d\Omega_{4-p}^2\right]
\ee
Now introducing a new coordinate by the relation $u^2 = r^{1-p}$ we can
rewrite the metric in \eqref{metricdpdp4} and the other field configurations
in \eqref{dpdp4} in terms of $u$ as,
\bea\label{dpdp4inu}
ds^2 &=& \sqrt{Q_1Q_2}u^2\left[-\frac{dt^2}{Q_1Q_2u^{\frac{4(2-p)}{1-p}}} +
  \frac{\sum_{i=1}^p (dx^i)^2}{Q_1Q_2u^{\frac{4(2-p)}{1-p}}} + 
\frac{\sum_{j=p+1}^{p+4} 
(dx^j)^2}{Q_2 u^2}+ \frac{4}{(1-p)^2}\frac{du^2}{u^2}\right.\nn
& & \left. \qquad\qquad\qquad\qquad\qquad\qquad\qquad\qquad\qquad\qquad\qquad 
\qquad\qquad\qquad+ d\Omega_{4-p}^2\right]\nn
e^{2\phi} &=& \frac{Q_1^{\frac{3-p}{2}}}{Q_2^{\frac{p+1}{2}}} u^{2(3-p)}\nn
A_{[p+1]} &=& -\frac{1}{Q_1u^{\frac{2(3-p)}{1-p}}}dt\wedge dx^1 
\wedge \ldots \wedge dx^p,
\quad A_{[p+5]}\,\,=\,\,-\frac{1}{Q_2u^{\frac{2(3-p)}{1-p}}} 
dt\wedge dx^1 \wedge \ldots 
\wedge dx^{p+4}\nn   
\eea   
It is clear from the metric in \eqref{dpdp4inu} that under the scaling $ u \to
\lambda u$, the coordinates $x^{1,\ldots,p}$ and $x^{p+1,\ldots,p+4}$ scale
differently if the part of the metric in square bracket has to remain
invariant. So, we will discuss $p=0$ and $p=2$ cases separately.

\subsection{$p=0$ or D0-D4 case}

In this case we observe from \eqref{dpdp4inu} that under the scaling $t \to 
\lambda^4 t \equiv \lambda^z t$, $x^{1,\ldots,4} \to \lambda x^{1,\ldots,4}$, 
$u \to \lambda u$, where $z$ in $t$ transformation is called the dynamical
critical exponent,
the metric in the square bracket is invariant. However, the full metric is not
invariant as there is a hyperscaling violation 
\cite{Charmousis:2010zz,Dong:2012se,Huijse:2011ef}. To find the hyperscaling
violation exponent we need to perform a dimensional reduction of the theory 
on S$^4$ and
express the resulting metric in Einstein frame. The reduced metric in this
case can be seen to transform as $ds_6 \to \lambda^{1/2} ds_6 \equiv
\lambda^{\theta/d} ds_6 \equiv \lambda^{\theta/4} ds_6$, where $\theta$ is the
hyperscaling violation exponent and $d$ is the spatial dimension of the
boundary theory. We thus find that the near horizon limit of intersecting 
D0-D4 solution has a Lifshitz-like metric with $z=4$ and $\theta=2$. The
dilaton and the form fields also transform (see \eqref{dpdp4inu}) under 
the above scaling as,
$\phi \to \phi + 3 \log \lambda$, $A_{[1]} \to \lambda^{-2} A_{[1]}$ and
$A_{[5]} \to \lambda^2 A_{[5]}$. It can be easily checked that the pair 
$(z,\theta)$ obtained in this case satisfy the null energy condition (NEC)
 \cite{Dong:2012se}
\bea\label{NEC}
(d-\theta)(d(z-1)-\theta) &\geq& 0\nn
(z-1)(d+z-\theta) &\geq& 0
\eea
As the dilaton is not constant, it will produce an RG flow in the boundary
theory as $u$ varies. However, as $u$ varies, the effective string coupling
$e^{\phi}$ and the curvature of the metric must remain small for the gravity
description to remain valid. This gives a restriction on $u$ as,
$1/(Q_1Q_2)^{1/4} \ll u \ll Q_2^{1/12}/Q_1^{1/4}$. But if $u \geq
Q_2^{1/12}/Q_1^{1/4}$ the dilaton becomes large and we have to uplift the
solution to M-theory. The eleven dimensional metric has the form,
\be\label{d0d411}
ds^2 = Q_2^{\frac{2}{3}}\left[-\frac{2}{Q_2u^2} dx^{11}dt + \frac{Q_1}{Q_2}
u^4 (dx^{11})^2 + \frac{\sum_{i=1}^4(dx^i)^2}{Q_2u^2} + 4\frac{du^2}{u^2}
+ d\Omega_4^2\right]
\ee
The above metric represents an intersecting solution of an M5 brane along
$x^1,\ldots,x^4,x^{11}$ with a wave along $x^{11}$. This gravity solution is
valid as long as $Q_2 \gg 1$. From \eqref{d0d411} we find that the metric is
invariant under an assymetric Lifshitz scaling $t \to \lambda^4 t$,
$x^{1,\ldots,4} \to \lambda x^{1,\ldots,4}$, $x^{11} \to \lambda^{-2} x^{11}$
and $u \to \lambda u$ without any hyperscaling violation ($\theta=0$).

\subsection{$p=2$ or D2-D6 case}

It can be seen from the metric in \eqref{dpdp4inu} that for $p=2$, the part of
the metric in the square bracket is invariant under the scaling $t \to 
\lambda^0 t$, $x^{3,4,5,6} \to \lambda x^{3,4,5,6}$ and $u \to \lambda u$. 
Note that the coordinates $x^{1,2}$ do not scale. However, the full metric is
not invariant as in $p=0$ case and therefore there is a hyperscaling
violation. In order to find its value we have to compactify the theory on 
S$^2$ $\times$ T$^2$ as the coordinates $x^{1,2}$ do not scale. The compact
theory will therefore be six dimensional and the spatial dimension of the
boundary theory is four. Expressing the compact metric in Einstein frame we
find that it transforms under the above scaling as $ds_6 \to
\lambda^{3/2} ds_6 \equiv \lambda^{\theta/d} ds_6$. We therefore find that the
near horizon limit of the intersecting D2-D6 solution has a Lifshitz-like
metric with the critical dynamical exponent $z=0$ and the hyperscaling
violation exponent $\theta=6$. This pair of $(z,\theta)$ can again be seen to
satisfy the NEC \eqref{NEC}. The dilaton and the form fields transform under
the scaling as $\phi \to \phi + \log\lambda$, $A_{[3]} \to \lambda^2 A_{[3]}$
and $A_{[7]} \to \lambda^6 A_{[7]}$. The dilaton is again found to be
non-constant and varies with $u$ giving an RG flow to the boundary theory.
But as the dilaton varies, the effective string coupling $e^{\phi}$ and the
curvature of the metric must remain small so that the gravity description can
be trusted. This gives a restriction on $u$ as, $1/(Q_1Q_2)^{1/4} \ll u \ll
Q_2^{3/4}/Q_1^{1/4}$. When $u \geq Q_2^{3/4}/Q_1^{1/4}$, the effective string
coupling becomes large and we have to uplift the theory to eleven dimensions.
The eleven dimensional metric in this case has the form,
\bea\label{d2d611}
ds^2 &=& Q_1^{1/3}Q_2 u^{4/3}\left[-\frac{dt^2}{Q_1Q_2}+ \frac{\sum_{i=1}^2
(dx^i)^2}{Q_1Q_2}+\frac{\sum_{j=3}^6 (dx^j)^2}{Q_2u^2} + 4\frac{du^2}{u^2}
+d\Omega_2^2\right.\nn
& &\left. \qquad\qquad\qquad\qquad\qquad\qquad\qquad\qquad\qquad
+ \frac{1}{Q_2^2}\left(dx^{11} - 2Q_2\sin^2(\theta/2)d\phi\right)^2\right]
\eea  
The solution \eqref{d2d611} represents an intersecting solution of M2 branes 
with
KK monopole. Under the scaling $t \to \lambda^0 t$, $x^{3,4,5,6} \to \lambda
x^{3,4,5,6}$ and $u\to \lambda u$, the part of the metric in the square
bracket remains invariant. Note that $x^{1,2}$ do not scale. Also, as the
whole metric is non-invariant there is a hyperscaling violation. We compactify
the theory on S$^2$ and also along $x^{1,2,11}$ to obtain the hyperscaling
violation exponent. Expressing the reduced metric in the Einstein frame we
find that it transforms under the above scaling as, $ds_6 \to \lambda^{3/2}
ds_6 \equiv \lambda^{\theta/d} ds_6$, where $d(=4)$ is the spatial dimension of
the boundary theory. We thus find that the eleven dimensional metric also has
a Lifshitz-like structure with dynamical critical exponent $z=0$ and
hyperscaling violation exponent $\theta=6$. This pair of $(z,\theta)$
also satisfies the NEC \eqref{NEC}. 
 
\section{D$p$-D$(p+2)$, D2-D2$'$, D3-D3$'$, D4-D4$'$ 
and Lifshitz-like metrics}
 
In \cite{Dey:2012tg}, we obtained intersecting D1-D3 solution of type 
IIB string theory
which is 1/4 BPS and threshold bound state. S-dual of this is the F-D3
solution discussed there. D-string in D1-D3 solution is transverse to the
D3-brane directions and are delocalized. If we apply T-duality to the common
transverse directions of D1-D3 solution we obtain D2-D4 and D3-D5 solution,
where D2-D4 intersects on a string and D3-D5 intersects on a membrane. We thus
obtain D$p$-D$(p+2)$ solutions for $p=1,\,2,\,3$. On the other hand if we 
apply T-duality along one of the D3-brane directions, we
obtain D2-D2$'$ bound state, where the two D2-branes are transverse 
(intersect on a point) to each
other. Further applying T-dualities along the common transverse directions of
D2-D2$'$ we can get D3-D3$'$ intersecting on a string and also D4-D4$'$
intersecting on a membrane. Some of these solutions were obtained in 
\cite{Gauntlett:1996pb,Behrndt:1996pm}.
We will show that all these solutions in the near 
horizon limit give Lifshitz-like metrics. We will also study their phase 
structures. Since the
different solutions have their own peculiarities, we can not study them in
generality and therefore discuss each case separately.

\subsection{D1-D3 case}

This case along with its S-dual version have already been discussed in 
\cite{Dey:2012tg}
and so we will not repeat it here. We found that both of them have
Lifshitz-like structures with $z=4$ and $\theta=2$.

\subsection{D2-D4 case} 

This configuration can be obtained by applying T-duality along, say, $x^5$ 
on the D1-D3 solution given in eq.(5) of ref.\cite{Dey:2012tg}. It has the 
form,
\bea\label{d2d4}
ds^2 &=& H_1^{\half}H_2^{\half}\left[-H_1^{-1}H_2^{-1}dt^2 + H_2^{-1}
\sum_{i=1}^3(dx^i)^2 + H_1^{-1}(dx^4)^2 + H_1^{-1}H_2^{-1}(dx^5)^2
\right.\nn
& & \Big.\qquad\qquad\qquad\qquad\qquad\qquad\qquad\qquad\qquad\qquad\qquad
 + dr^2
+ r^2d\Omega_3^2\Big]\nn
e^{2\phi} &=& \left(\frac{H_1}{H_2}\right)^{\half}\nn
A_{[3]} &=& (1-H_1^{-1})dt\wedge dx^4\wedge dx^5, \quad A_{[5]}\,\,=\,\,
(1-H_2^{-1})dt\wedge dx^1\wedge dx^2\wedge dx^3 \wedge dx^5
\eea
The harmonic functions in \eqref{d2d4} are given as, $H_{1,2} = 1+ 
Q_{1,2}/r^2$, with the radial coordinate $r = \sqrt{(x^6)^2+\cdots +
  (x^9)^2}$. 
It is clear
from the metric that D2 branes lie along $x^4$, $x^5$, whereas D4 branes
lie along $x^1$, $x^2$, $x^3$, $x^5$. In the near horizon limit we
approximate $H_{1,2} \approx Q_{1,2}/r^2$ and then making the coordinate
change $r \to 1/r$, the above configuration \eqref{d2d4} takes the form,
\bea\label{d2d4NH}
ds^2 &=& \sqrt{Q_1Q_2}\left[-\frac{dt^2}{Q_1Q_2r^2} + \frac{\sum_{i=1}^3
(dx^i)^2}{Q_2} + \frac{(dx^4)^2}{Q_1}+\frac{(dx^5)^2}{Q_1Q_2r^2} +
\frac{dr^2}{r^2} + d\Omega_3^2\right]\nn
e^{2\phi} &=& \left(\frac{Q_1}{Q_2}\right)^{\half}\nn
A_{[3]} &=& -\frac{1}{Q_1r^2} dt \wedge dx^4 \wedge dx^5,\quad
A_{[5]}\,\, =\,\, -\frac{1}{Q_2r^2} dt \wedge dx^1 \wedge dx^2 \wedge dx^3 
\wedge dx^5
\eea
We thus find that D2-D4 solution does not give Lifshitz-like scaling
in the near horizon limit, rather it gives an AdS$_3$ space. The gravity
description in this case is valid for $1/Q_2 \ll Q_1 \ll Q_2$. However,
if $Q_1>Q_2$, the effective string coupling would be large and we have to
uplift the theory to eleven dimensions. The eleven dimensional metric
has the form,
\be\label{d2d411}
ds^2 = Q_1^{\frac{1}{3}} Q_2^{\frac{2}{3}}\left[-\frac{dt^2}{Q_1Q_2r^2}
+ \frac{\sum_{i=1}^3(dx^i)^2}{Q_2} + \frac{(dx^4)^2}{Q_1} + 
\frac{(dx^5)^2}{Q_1Q_2r^2} + \frac{(dx^{11})^2}{Q_2} + \frac{dr^2}{r^2}
+ d\Omega_3^2\right]
\ee        
The above solution represents the near horizon limit of intersecting
M2-M5 branes meeting on a string where M2 branes are along $x^4$ and
$x^5$, and M5 branes are along $x^{1,2,3}$, $x^5$ and $x^{11}$.
As it is clear this uplifted solution also has AdS$_3$ structure. This gravity 
description can be trusted as long as $Q_1\gg 1/Q_2^2$.
 
\subsection{D3-D5 case}

This state can be constructed by applying T-duality along one of the common
transverse directions ($x^6$ say) of D2-D4 solution given in \eqref{d2d4}.
The solution takes the form,
\bea\label{d3d5}
ds^2 &=& H_1^{\half}H_2^{\half}\left[-H_1^{-1}H_2^{-1}dt^2 + H_2^{-1}
\sum_{i=1}^3(dx^i)^2 + H_1^{-1}(dx^4)^2 + H_1^{-1}H_2^{-1}((dx^5)^2
+(dx^6)^2)
\right.\nn
& & \Big.\qquad\qquad\qquad\qquad\qquad\qquad\qquad\qquad\qquad\qquad\qquad
 + dr^2
+ r^2d\Omega_2^2\Big]\nn
e^{2\phi} &=& \frac{1}{H_2}, \qquad F_{[5]} \,\,=\,\, 
-(1+\ast)dH^{-1} \wedge dt\wedge dx^4 \wedge \ldots \wedge dx^6\nn 
A_{[6]} &=&
(1-H_2^{-1})dt\wedge dx^1\wedge \ldots \wedge dx^3 \wedge dx^5\wedge dx^6
\eea
The harmonic functions in this case are given as $H_{1,2} =
1+Q_{1,2}/r$. Here D3-branes lie along $x^4$, $x^5$, $x^6$ and D5-branes lie
along $x^1$, $x^2$, $x^3$, $x^5$, $x^6$. Also note that in the above we have 
given the form of the field-strength (instead of the gauge field) as it is 
self-dual.

Now taking the near horizon limit $H_{1,2} \approx Q_{1,2}/r$, then changing
the corrdinate $r$ by $r \to 1/r$ and finally, introducing a new coordinate by
$u^2=1/r$, we can rewrite the D3-D5 configuration in the near horizon limit
as,
\bea\label{d3d5inu}
ds^2 &=& \sqrt{Q_1Q_2}u^2\left[-\frac{dt^2}{Q_1Q_2}+\frac{\sum_{i=1}^3(dx^i)^2}
{Q_2u^2}+\frac{(dx^4)^2}{Q_1u^2}+\frac{(dx^5)^2+(dx^6)^2}{Q_1Q_2}+4
\frac{du^2}{u^2} + d\Omega_2^2\right]\nn
e^{2\phi} &=& \frac{u^2}{Q_2},\qquad F_{[5]}\,\,=\,\, -(1+\ast)\frac{2u}{Q_1}
du\wedge dt \wedge dx^4\wedge dx^5\wedge dx^6\nn
A_{[6]} &=& -\frac{u^2}{Q_2} dt \wedge dx^1\wedge dx^2\wedge dx^3\wedge dx^5
\wedge dx^6
\eea
We find from \eqref{d3d5inu} that under the scaling $t \to \lambda^0 t$, 
$x^{1,2,3,4} \to \lambda x^{1,2,3,4}$, $u \to \lambda u$, the part of the
metric in the square bracket is invariant, but the full metric is
not and so there is a hyperscaling violation. Note that $x^{5,6}$
do not scale. To find the hyperscaling violation exponent ($\theta$) 
we have to 
compactify the theory on S$^2$ and also on $x^{5,6}$ and then express
the resulting metric in the Einstein frame. This way we find that 
the reduced Einstein frame metric transforms under the above scaling as,
$ds_6 \to \lambda^{3/2} ds_6 \equiv \lambda^{\theta/d} ds_6$, where
$d=4$. Therefore, we get $\theta=6$. We thus find that D3-D5 solution in the
near horizon limit has a Lifshitz-like metric with a dynamical critical
exponent $z=0$ and a hyperscaling violation exponent $\theta=6$. This pair can
be shown to satisfy NEC \eqref{NEC}. The dilaton and the form fields can be
shown to transform under the above scaling as, $\phi \to \phi + \log \lambda$,
$F_{[5]} \to \lambda^3 F_{[5]}$ and $A_{[6]} \to \lambda^5 A_{[6]}$.

The gravity description \eqref{d3d5inu} is valid if $u$ lies in the range
$1/(Q_1Q_2)^{1/4} \ll u \ll Q_2^{1/2}$. However for $u \geq Q_2^{1/2}$, the
effective string coupling $e^{\phi}$ becomes large and the gravity description 
breaks down. For that we have to go to the S-dual frame where the metric and
the other fields take the form,    
\bea\label{d3d5inuSdual}
ds^2 &=& Q_1^{\half}Q_2u\left[-\frac{dt^2}{Q_1Q_2}+\frac{\sum_{i=1}^3(dx^i)^2}
{Q_2u^2}+\frac{(dx^4)^2}{Q_1u^2}+\frac{(dx^5)^2+(dx^6)^2}{Q_1Q_2}+4
\frac{du^2}{u^2} + d\Omega_2^2\right]\nn
e^{2\phi} &=& \frac{Q_2}{u^2},\qquad F_{[5]}\,\,=\,\, -(1+\ast)\frac{2u}{Q_1}
du\wedge dt \wedge dx^4\wedge dx^5\wedge dx^6\nn
H_{[3]} &=& -Q_2 dx^4 \wedge \epsilon_2
\eea
This represents the near horizon limit of intersecting 1/4 BPS D3-NS5 
threshold bound state solution. Under the same scaling as in D3-D5 we find 
that the part of the metric (see \eqref{d3d5inuSdual}) in square bracket is
invariant. However from the transformation of the reduced metric we find that
it has a Lifshitz-like structure with the same $(z,\theta)=(0,6)$ as in the
D3-D5 case. Here the other fields transform as, $\phi \to \phi - \log\lambda$,
$F_{[5]} \to \lambda^3 F_{[5]}$ and $H_{[3]} \to \lambda H_{[3]}$.

\subsection{D2-D2$'$ case}

As mentioned earlier,
D2-D2$'$ intersecting solution can be obtained by applying T-duality 
along one of the D3 brane directions (say, $x^3$) of the
D1-D3 solution given in eq.(5) of ref.\cite{Dey:2012tg}. It has the form,
\bea\label{d2d2prime}
ds^2 &=& H_1^{\half}H_2^{\half}\left[-H_1^{-1}H_2^{-1}dt^2 + H_2^{-1}
\sum_{i=1}^2(dx^i)^2 + H_1^{-1}\sum_{j=3}^4(dx^j)^2 
 + dr^2 + r^2d\Omega_4^2\right]\nn
e^{2\phi} &=& (H_1 H_2)^{\half}\nn
A_{[3]} &=& (1-H_1^{-1})dt\wedge dx^4\wedge dx^3, \quad A_{[3]}'\,\,=\,\,
(1-H_2^{-1})dt\wedge dx^1\wedge dx^2
\eea
Here the harmonic functions are given as $H_{1,2}=1+Q_{1,2}/r^3$. The two D2
branes are along $x^1$, $x^2$ and $x^3$, $x^4$. Going to the near horizon
limit $H_{1,2} \approx Q_{1,2}/r^3$, changing from $r \to 1/r$ and introducing
 a new coordinate by $u^2=r$, we obtain from \eqref{d2d2prime}
\bea\label{d2d2primeinu}
ds^2 &=& (Q_1Q_2)^{\half} u^2\left[-\frac{dt^2}{Q_1Q_2u^8} +
\frac{\sum_{i=1}^2(dx^i)^2}{Q_2u^2} + \frac{\sum_{j=3}^4(dx^j)^2}{Q_1u^2} 
 + 4\frac{du^2}{u^2} + d\Omega_4^2\right]\nn
e^{2\phi} &=& (Q_1Q_2)^{\half} u^6\nn
A_{[3]} &=& -\frac{1}{Q_1u^6}dt\wedge dx^4\wedge dx^3, \quad A_{[3]}'\,\,=\,\,
-\frac{1}{Q_2u^6}dt\wedge dx^1\wedge dx^2
\eea
Thus we find that under the scaling $t \to \lambda^4 t$, $x^{1,2,3,4} \to
\lambda x^{1,2,3,4}$ and $u \to \lambda u$, the part of the metric (given in
\eqref{d2d2primeinu}) in the square bracket remains invariant. But the whole
metric is not invariant because of the hyperscaling violation. As before, 
we find that the
reduced metric in the Einstein frame transforms under the above scaling as
$ds_6 \to \lambda^{1/2} ds_6 \equiv \lambda^{\theta/d} ds_6$. We thus find
$\theta=2$. Therefore, D2-D2$'$ solution in the near horizon limit has
Lifshitz-like metric with $z=4$ and $\theta=2$. This pair of $(z,\theta)$
can be shown to satisfy NEC \eqref{NEC}. The dilaton 
transforms under the scaling as, $\phi \to \phi + 3\log\lambda$. $A_{[3]}$ and
$A_{[3]}'$ remain invariant.

In this case the above gravity description is valid for $1/(Q_1Q_2)^{1/4} \ll
u \ll 1/(Q_1Q_2)^{1/12}$. When $u\geq 1/(Q_1Q_2)^{1/12}$, the effective string
coupling $e^{\phi}$ becomes large and the gravity description breaks down. In
that case we have to uplift the solution to M-theory. The uplifted solution
has the form,
\be\label{d2d2prime11}
ds^2 = (Q_1Q_2)^{\frac{1}{3}} u^{\frac{2}{3}} 
\left[-\frac{dt^2}{Q_1Q_2 u^6}
+ \frac{\sum_{i=1}^2 (dx^i)^2}{Q_2 u^2} + 
\frac{\sum_{j=3}^4 (dx^j)^2}{Q_1 u^2} + \frac{du^2}{u^2} + 
d\Omega_{5}^2\right]
\ee
This represents two intersecting M2 branes along $x^1,\,
x^2$ and $x^3,\,x^4$. The part of the metric in the 
square bracket has the scale invariance $t \to \lambda^3 t$,
$x^{1,2,3,4} \to \lambda x^{1,2,3,4}$, $u \to \lambda u$.
The metric has a Lifshitz-like structure with $(z,\,\theta)
= (3,\,3)$.

\subsection{D3-D3$'$ case}

This bound state can be obtained by applying T-duality along one of the common
transverse directions ($x^5$, say) of the D2-D2$'$ solution given in 
\eqref{d2d2prime}. This way we obtain D3-D3$'$ solution in the following form,
\bea\label{d3d3prime}
ds^2 &=& H_1^{\half}H_2^{\half}\left[-H_1^{-1}H_2^{-1}dt^2 + H_2^{-1}
\sum_{i=1}^2(dx^i)^2 + H_1^{-1}\sum_{j=3}^4(dx^j)^2 + H_1^{-1}H_2^{-1}
(dx^5)^2\right.\nn 
& &
\Big.\qquad\qquad\qquad\qquad\qquad\qquad\qquad\qquad\qquad\qquad\qquad\qquad  
+ dr^2 + r^2d\Omega_3^2\Big]\nn
e^{2\phi} &=& 1\nn
F_{[5]} &=& -(1+\ast)dH_1^{-1}\wedge dt\wedge dx^4\wedge dx^3 \wedge dx^5\nn 
F_{[5]}' &=&
-(1+\ast) dH_2^{-1}\wedge dt\wedge dx^1\wedge dx^2\wedge dx^5
\eea
Here the two D3-branes are along the directions $x^1$, $x^2$, $x^5$ and 
$x^3$, $x^4$, $x^5$,
i.e., they intersect on a string. The harmonic functions are given as $H_{1,2} =
1+Q_{1,2}/r^2$. $F_{[5]}$ and $F_{[5]}'$ are the two self-dual field-strengths
to which the D3-branes couple. In the near horizon limit $H \approx
Q_{1,2}/r^2$, along with the change of coordinates $r \to 1/r$, the 
configuration \eqref{d3d3prime} takes the form,    
\bea\label{d3d3primeNH}
ds^2 &=& (Q_1Q_2)^{\half}\left[-\frac{dt^2}{Q_1Q_2r^2} +
\frac{\sum_{i=1}^2(dx^i)^2}{Q_2} + \frac{\sum_{j=3}^4(dx^j)^2}{Q_1} +
\frac{(dx^5)^2}{Q_1Q_2r^2} 
 + \frac{dr^2}{r^2} + d\Omega_3^2\right]\nn
e^{2\phi} &=& 1\nn
F_{[5]} &=& (1+\ast) \frac{2}{Q_1r^3} dr\wedge dt\wedge dx^4\wedge dx^3 \wedge
dx^5\nn 
F_{[5]}' &=&
(1+\ast)\frac{2}{Q_2r^3}dr\wedge dt\wedge dx^1\wedge dx^2\wedge dx^5
\eea
It is clear from the above, that D3-D3$'$ indeed has AdS$_3$
structure in the near horizon limit. Note that the supergravity description is
valid as long as the string coupling $g_s$ (which is suppressed here) is small
and $Q_1 \gg 1/Q_2$.

\subsection{D4-D4$'$ case}

The application of a further T-duality along a common tansverse direction 
($x^6$, say) of D3-D3$'$ solution given in \eqref{d3d3prime} will produce 
D4-D4$'$ solution given as,
\bea\label{d4d4prime}
ds^2 &=& H_1^{\half}H_2^{\half}\left[-H_1^{-1}H_2^{-1}dt^2 + H_2^{-1}
\sum_{i=1}^2(dx^i)^2 + H_1^{-1}\sum_{j=3}^4(dx^j)^2 \right.\nn
& & \left. \qquad\qquad\qquad\qquad\qquad\qquad\qquad\qquad + H_1^{-1}H_2^{-1}
\sum_{k=5}^6(dx^k)^2   
+ dr^2 + r^2d\Omega_2^2\right]\nn
e^{2\phi} &=& \left(H_1H_2\right)^{-\half}\nn
A_{[5]} &=& (1-H_1^{-1}) dt\wedge dx^4\wedge dx^3 \wedge dx^5 \wedge dx^6\nn 
A_{[5]}' &=&
(1-H_2^{-1}) dt\wedge dx^1\wedge dx^2\wedge dx^5 \wedge dx^6
\eea
From \eqref{d4d4prime} we observe that the two D4-branes in this solution lie
along $x^1$, $x^2$, $x^5$, $x^6$ and $x^3$, $x^4$, $x^5$, $x^6$, i.e., they
intersect on a membrane. The harmonic functions are given as $H_{1,2} =
1+ Q_{1,2}/r$. Taking the near horizon limit $H_{1,2} \approx Q_{1,2}/r$,
changing coordinates $r \to 1/r$ and introducing new variable by $u^2 = 1/r$,
we rewrite the above solution in terms of this new variable as,
\bea\label{d4d4primeinu}
ds^2 &=& (Q_1Q_2)^{\half}u^2\left[-\frac{dt^2}{Q_1Q_2} +
\frac{\sum_{i=1}^2(dx^i)^2}{Q_2u^2} + \frac{\sum_{j=3}^4(dx^j)^2}{Q_1u^2} +
\frac{\sum_{k=5}^6(dx^k)^2}{Q_1Q_2} 
 + 4\frac{du^2}{u^2} + d\Omega_2^2\right]\nn
e^{2\phi} &=& \frac{u^2}{(Q_1Q_2)^{\half}}\nn
A_{[5]} &=& -\frac{u^2}{Q_1} dt\wedge dx^4\wedge dx^3 \wedge
dx^5 \wedge dx^6\nn 
A_{[5]}' &=&
-\frac{u^2}{Q_2}dt\wedge dx^1\wedge dx^2\wedge dx^5\wedge dx^6
\eea 
We observe from \eqref{d4d4primeinu} that the part of the metric in square
bracket is invariant under the scaling $t \to \lambda^0 t$, $x^{1,2,3,4} \to 
\lambda x^{1,2,3,4}$, $u \to \lambda u$. However, the full metric is not
invariant under this scaling and so there is a hyperscaling
violation. Compactifying the theory on S$^2$ and also along $x^{5,6}$ (as they
do not scale) we find that the reduced metric in Einstein frame transforms as
$ds_6 \to \lambda^{3/2} ds_6 \equiv \lambda^{\theta/d} ds_6$, where $d(=4)$ is
the spatial dimension of the boundary theory. This gives $\theta=6$. We 
therefore find that D4-D4$'$
solution in the near horizon limit has a Lifshitz-like metric with $z=0$ and
$\theta=6$. As before this pair $(z=0,\,\theta=6)$ satisfies NEC \eqref{NEC}.
The dilaton and the form fields transform under the above scaling
as $\phi \to \phi + \log\lambda$, $A_{[5]} \to \lambda^4 A_{[5]}$ and  
$A_{[5]}' \to \lambda^4 A_{[5]}'$. 

We note that the above gravity description is valid when the effective string
coupling $e^{\phi}$ and the curvature of the metric remain small. This gives
the restriction on $u$ as $1/(Q_1Q_2)^{1/4} \ll u \ll (Q_1Q_2)^{1/4}$. However
when $u \geq (Q_1Q_2)^{1/4}$ we have to uplift the solution to M-theory. The
eleven dimensional solution can be seen to take the form,
\bea
ds^2 &=& (Q_1Q_2)^{\frac{2}{3}}u^{\frac{4}{3}}\left[-\frac{dt^2}{Q_1Q_2} +
\frac{\sum_{i=1}^2(dx^i)^2}{Q_2u^2} + \frac{\sum_{j=3}^4(dx^j)^2}{Q_1u^2} +
\frac{\sum_{k=5}^6(dx^k)^2}{Q_1Q_2} 
 + 4\frac{du^2}{u^2} + d\Omega_2^2 \right.\nn
& & \Big. \qquad\qquad\qquad\qquad\qquad\qquad\qquad\qquad\qquad\qquad\qquad
\qquad\qquad 
+ \frac{(dx^{11})^2}{Q_1Q_2}\Big]\nn
A_{[6]} &=& -\frac{u^2}{Q_1} dt\wedge dx^4\wedge dx^3 \wedge
dx^5 \wedge dx^6 \wedge dx^{11}\nn 
A_{[6]}' &=&
-\frac{u^2}{Q_2}dt\wedge dx^1\wedge dx^2\wedge dx^5\wedge dx^6\wedge dx^{11}
\eea 
This solution represents two M5 branes intersecting on a three brane along
$x^5$, $x^6$ and $x^{11}$. We again find that the metric has Lifshitz-like
scaling as the part of the metric in the square bracket is invariant under
same scaling as the D4-D4$'$ solution. Again as the full metric is not scale 
invariant, there is a hyperscaling violation. The hyperscaling violation
exponent can be found as before by reducing the metric on S$^2$ and also
on $x^{5,6,11}$ and expressing the resulting metric in Einstein frame. We thus
find that $\theta$ has the value 6. Therefore, D4-D4$'$ solution in the strong
coupling phase also has Lifshitz-like scaling with $z=0$ and $\theta=6$.

\section{Conclusion}

To conclude, in this paper we have shown how Lifshitz-like metrics (space-time
metrics having Lifshitz scaling with hyperscaling violation) arise from
the near horizon limit of certain intersecting D-brane solutions of type II
string theories. Some of these solutions are standard 1/4 BPS D0-D4, D2-D6
threshold bound states and some are non-standard 1/4 BPS D1-D3, D2-D4, D3-D5
along with D2-D2$'$, D3-D3$'$, D4-D4$'$ threshold bound states. All these 
solutions (except D2-D4 and D3-D3$'$) in the near horizon limit gave rise to
Lifshitz-like metrics with some dynamical critical exponent $z$ and some
hyperscaling violation exponent $\theta$. D2-D4 and D3-D3$'$ solutions gave 
AdS$_3$
spaces. We found that in all these solutions except D2-D4 and D3-D3$'$, 
the dilatons were
non-constant. We discussed also the phase structures of various solutions.
The metrics in other phases were also found to have Lifshitz-like structures.
For the various solutions, we found that there are two sets of values for the
dynamical critical exponent ($z$), the hyperscaling violation exponent
($\theta$) and the 
spatial dimension ($d$) of the boundary theory. The solutions D2-D6, D3-D5
and D4-D4$'$ 
as well as their strongly coupled phases yielded Lifshitz-like metrics in the 
near horizon limit with $(z=0,\,\theta=6,\, d=4)$. On the other hand, the
solutions D0-D4, D1-D3 (and its strongly coupled phase) and D2-D2$'$  
yielded Lifshitz-like metrics in the near horizon
limit with $(z=4,\,\theta=2,\,d=4)$. We have checked that both these values
satisfy null energy condition. However, none of these values satisfy
$\theta=d-1$. As emphasized in \cite{Ogawa:2011bz,Huijse:2011ef} that 
theories of this type may be of 
interest to some condensed matter system such as 
compressible metallic states with hidden Fermi surface. Whether theories
corresponding to the values of $z$ and $\theta$ we have obtained in this paper
have any potential application to condensed matter system is yet to be seen.
           
\section*{Acknowledgements}

One of the authors (PD) would like to acknowledge thankfully the financial
support of the Council of Scientific and Industrial Research, India
(SPM-07/489 (0089)/2010-EMR-I).

\end{document}